\documentclass[utf8]{FrontiersinHarvard} 
\usepackage{url,hyperref,subcaption}

\usepackage[nopatch=footnote]{microtype}

\usepackage[onehalfspacing]{setspace}

\usepackage{graphicx}
\usepackage{caption}
\usepackage{amssymb}
\usepackage{lineno}

\def\keyFont{\fontsize{8}{11}\helveticabold }
\def\firstAuthorLast{Gupta {et~al.}} 
\def\Authors{Sakshi Gupta\,$^{1,2,3}$, Arnab Basak\,$^{1,4}$ and Dibyendu Nandy\,$^{1,2}$}
\def\Address{$^{1}$Center of Excellence in Space Sciences India, Indian Institute of Science Education and Research Kolkata, Mohanpur 741246, India \\
$^{2}$Department of Physical Sciences, Indian Institute of Science Education and Research Kolkata, Mohanpur 741246, India \\
$^{3}$Center for High Energy Systems and Sciences, Defence Research and Development Organization, Hyderabad, 500058, India \\
$^{4}$Department of Physics, Fakir Chand College, Diamond Harbour 743331, India}
\def\corrAuthor{Dibyendu Nandy}

\def\corrEmail{dnandi@iiserkol.ac.in}

\DeclareUnicodeCharacter{22C6}{$\star$}

\begin{document}
\onecolumn
\firstpage{1}

\title[Alfvén Wing and Magnetospheric Dynamics]{Stellar Wind Driven Alfvén Wing Dynamics in Planetary and Exoplanetary Magnetospheres}

\author[\firstAuthorLast ]{\Authors} 
\address{} 
\correspondance{} 

\extraAuth{}

\maketitle

\begin{abstract}

Magnetized obstacles embedded within a plasma flow generate magnetohydrodynamic structures known as Alfvén wings, which act as primary conduits for the transfer of momentum and energy between the body and the surrounding medium. This study employs three-dimensional resistive magnetohydrodynamic simulations to explore how these wings and the magnetosphere respond to diverse stellar wind conditions. Our results, gleaned from a large number of systematic simulations spanning a wide range of stellar wind speed and magnetic field -- and planetary dipole field -- show that the global magnetospheric configuration is highly sensitive to the upstream Alfvén Mach number. We find that increasing stellar wind speed leads to a systematic closure and narrowing of Alfvén wing structures, while stronger stellar magnetic fields facilitate their opening. Analysis of the Alfvén wing morphology demonstrates a distinct dependence of wing opening angle on stellar wind speed, with internal wing analysis showing a reduction in plasma velocity and significant magnetic-flux accumulation. Our results exhibit a clear interdependence between the day-side magnetopause stand-off distance and the night-side magnetotail current sheet length. We find a linear scaling between the magnetotail dynamics and upstream forcing parameters. This study bridges the gap between solar system observations and (exo)planetary systems by demonstrating how Earth-like magnetospheres might transform into wing-dominated configurations during extreme stellar events or within the sub-Alfvénic regimes of close-in (exo)planets. Our findings can aid the interpretation of Alfvén wing signatures in observational data and enhance our understanding of how (exo)planetary magnetospheres respond to dynamic stellar wind forcing.

\tiny
 \keyFont{ \section{Keywords:} Star-planet interaction, Alfven wing, magnetohydrodynamics, (exo)planetary magnetospheric dynamics, stellar winds, planetary magnetosphere} 
\end{abstract}

\section{Introduction}

The in-situ detection of Alfvén wings at Earth during the May 2002 and April 2023 solar wind conditions \citep{Burkholder_2024, chen_2024} has evidenced that our own planet can undergo a large magnetospheric transformation when the solar wind transitions to sub-Alfvénic regime. Such conditions are expected to be routine for (exo)planets orbiting magnetically active stars or for the close-in (exo)planets . Magnetospheres act as dynamic shields that regulate how stellar winds interact with planets, controlling the transfer of energy, mass, and momentum between the star and the planetary environment. Rather than being static cavities, magnetospheres continually expand or contract in response to variations in stellar wind dynamic pressure. Depending on orbital distance, planetary and stellar wind properties, planets can experience very different wind regimes, which in turn lead to distinct magnetospheric configurations i.e. variations in magnetopause standoff distance, bow-shock structure, and magnetotail dynamics (\citep{Lammer_2007,Vidotto_2014,Cohen_2014,basak_nandy_2021, Gupta_2023}

The Alfvén Mach number $M_{A}$, defined as the ratio of the plasma flow speed to the local Alfvén speed, is a fundamental controlling parameter of plasma behavior in magnetized environments. Typically, the solar wind has $M_{A}$
values around 8 or higher, indicating that the flow speed greatly exceeds the Alfvén speed. However, during transient solar events such as coronal mass ejections (CMEs), $M_{A}$ can drop significantly, altering the interaction with planetary magnetospheres \citep{Ridley_2007, Burkholder_2024}. 

Stellar wind parameters such as speed and magnetic field strength are highly variable across stellar types and evolutionary stages, with young, magnetically active stars producing winds that can be orders of magnitude denser and more magnetized than the present-day solar wind at Earth \citep{ NANDY2007891,Vidotto_2014, vidotto_2020}. These conditions are especially relevant for close-in exoplanets, which can frequently encounter sub- or near-Alfvénic stellar winds \citep{Cohen_2014}. Highly active M dwarfs and young solar analogues can expose their planets to sustained intervals of low $M_{A}$, reshaping planetary magnetospheres and modulating star-planet interactions \citep{see_2014, Garraffo_2017,Airapetian_2020} . Understanding how planetary magnetospheres respond under low and variable Alfvén Mach number regimes is therefore critical for interpreting planetary magnetospheric dynamics, assessing space weather phenomena and predicting electromagnetic signatures across diverse stellar systems.

One of the features that emerges under sub-Alfvénic plasma flow is the formation of extended Alfvén wings \citep{drell_1965,Neubauer_1980}. These magnetohydrodynamic structures — comprising current sheets aligned along perturbed magnetic field lines — arise when the plasma flow speed is slower than the local Alfvén speed, allowing perturbations to propagate upstream and establish standing wave patterns. Analytically, the wing opening angle is predicted to scale as $\theta_A = \arctan(1/M_A)$ for a point-source conducting obstacle \citep{Neubauer_1980, saur_2013}. Alfvén wings have been observed at solar system bodies including Io and Ganymede within Jupiter's magnetosphere 
\citep{Kivelson_1997,jia_2008}, at Mercury during a sub-Alfvénic interplanetary CME event \citep{Bowers_2025} and transiently at Earth during extreme solar events \citep{chane_2012,Burkholder_2024,chen_2024,Gurram_2025,Beedle_2024}. Global MHD simulations have reproduced these wing structures and demonstrated that their morphology and opening angle are sensitive to both wind speed and ambient field strength \citep{Ridley_2007, jia_2008, chane_2012, Corso_2026}. Observational signatures of these wings can be helpful in analyzing the star-planet magnetic interaction \citep{Nichols_2016, Szalay_2020,Pineda2023}. 

The magnetotail current sheet links dayside and nightside magnetospheric dynamics, and its structure directly reflects the upstream stellar wind conditions driving the system \citep{Dungey_1961,Slavin_1985,Lockwood_1995}. Within the Dungey-cycle framework, dayside reconnection opens magnetic flux that is transported into the tail lobes and later closed by nightside reconnection, so that changes in stellar wind conditions simultaneously affect both the magnetopause standoff distance and the tail geometry \citep{Dungey_1961, Milan_2021}. Stronger driving produces a more stretched tail with a longer current sheet favoring enhanced reconnection and plasmoid release, while weaker driving allows a shorter, more dipolar configuration \citep{Sergeev_2012, Lugaz_2016, Ganushkina_2018}. These relationships have been extended to exoplanetary magnetospheres through global MHD simulations exploring how stellar wind conditions influence magnetopause standoff distances, bow-shock locations, and large-scale field topology (\citep{Arridge_2009,Vidotto_2014, Cohen_2014,Carolan_2019,basak_nandy_2021,Gupta_2023,Roy_2023} The degree to which Alfvén wing structures further influence atmospheric escape, ionospheric coupling, and long-term planetary habitability has also been explored for planets orbiting M-dwarf stars where extreme and sustained wind conditions are common \citep{Cohen_2014,vidotto_2020,Torres_2021,presa_2024}.

Despite these advances, most studies either address specific case studies or restricted regions of parameter space. The interplay between stellar wind speed, stellar magnetic field strength, and planetary dipole moment in shaping both Alfvén wing morphology and magnetotail structure has not been explored in a unified way across a broad parameter range. \cite{Gupta_2023} shows how varying stellar and planetary magnetic field strengths shape magnetospheric topology and drive atmospheric mass loss, providing a comprehensive understanding on magnetic field-driven magnetospheric response. 
Building on this foundation, the present study extends the parameter space to include stellar wind speed as a primary driver and focuses on characterizing Alfvén wing morphology and the scaling of magnetotail structure with upstream forcing parameters.

In this work, we present 90 three-dimensional resistive MHD simulations spanning a wide parameter space of stellar wind speeds ($1-10,V_{sw}$), stellar magnetic fields (10–50 nT), and planetary dipole strengths ($0.5-2,B_e$). We systematically characterize the morphological evolution of Alfvén wings from broad, open structures at low $M_A$ to collimated, flow-aligned geometries at high $M_A$, and validate the simulated wing opening angles with the analytical prediction of \citep{Neubauer_1980}. We discuss the relationships linking magnetopause standoff distance to magnetotail current sheet length, analyze the internal velocity and magnetic-field structure within the wings, and identify the scaling between the composite upstream forcing parameter $V_{sw}^2/B_{sw}$ and the magnetotail current sheet length. These results  provide a coherent framework for interpreting Alfvén wing signatures and magnetospheric
response across both solar system and exoplanetary environments subject to diverse stellar wind forcing.

This paper begins with the model description in section 2 in which we give an overview of the theory and numerical setup employed in this study. The detailed findings are presented in section 3 followed by the conclusion in section 4.

\section{Model Description}

We adapt the Star-Planet Interaction Module (CESSI-SPIM) developed by \cite{Das_2019}, based on the PLUTO MHD solver \citep{Mignone2007}, to simulate different configurations of stellar winds and planetary magnetospheres. The governing set of resistive MHD equations are given by:

\begin{equation}
  \partial_{t} \rho+\nabla \cdot(\rho \vec{v})=0  
\end{equation}
  
\begin{equation}
   \partial_{t} \vec{v}+(\vec{v} \cdot \nabla) \vec{v}+\frac{1}{4 \pi \rho} \vec{B} \times(\nabla \times \vec{B})+\frac{1}{\rho} \nabla P=\vec{g}
\end{equation}

\begin{equation}
   \partial_{t} E+\nabla \cdot[(E+P) \vec{v}-\vec{B}(\vec{v} \cdot \vec{B})+(\eta \cdot \vec{J}) \times \vec{B}]=\rho \vec{v} \cdot \vec{g}
\end{equation}

\begin{equation}
   \partial_{t} \vec{B}+\nabla \times(\vec{B} \times \vec{v})+\nabla \times(\eta \cdot \vec{J})=0
\end{equation}

where the symbols $\rho$, $v$, $\vec{B}$, $P$, $E$, and $\vec{g}$ denote density, velocity, magnetic field, pressure, total energy density and gravitational acceleration due to the planet respectively. The quantity $\vec{J}$ is the current density given by $\nabla \times \vec{B}$, ignoring the displacement current. The expression for total energy density is given by

\begin{equation}
    E = \frac{P}{\gamma-1} + \frac{\rho v^2}{2} + \frac{B^2}{8\pi}
\end{equation}
for an ideal gas equation of state.

The computational domain extends from -80 ${R_p}$ to 200 ${R_p}$ in the $x$-direction, -45 ${R_p}$ to 45 ${R_p}$ in the $y$-direction and -200 ${R_p}$ to 200 ${R_p}$ in the $z$-direction, where ${R_p}$ is the radius of the planet which is located at the origin of the Cartesian box. Considering the constraints of our modest computational facility, a combination of stretched and uniform grid types is used in the $x$ and $z$ directions for all the simulations in this study. We adopt a similar simulation set-up as studied in \cite{Gupta_2023}, in which we explore the impact of changing stellar and planetary magnetic field strength on the planetary magnetosphere.
For model description and grid resolution kindly refer to  \cite{Gupta_2023}. We consider a finite and isotropic magnetic diffusivity $\eta$, as the causal mechanism for non-ideal processes such as magnetic reconnections. The magnetic diffusivity lies within the range of realistic values as shown in earlier studies \citep{Raeder_1999, Komar_2013,Varela_2022}.

In this study, an Earth-like planet is considered with similar mass, radius, and tilt while the stellar wind properties such as magnetic field and velocity are varied. While the setup uses Earth-like parameters (mass, radius, tilt) for the planet, the varying $M_A$ allows these results to be scaled to different magnetized (exo)planets. We perform a comprehensive parameter-space study consisting of ninety independent three-dimensional simulations. These runs encompass all relevant combinations of the varying stellar wind speeds and magnetic fields as well as planetary dipole fields, allowing for a characterization of magnetospheric response across a wide parameter space and are necessary for understanding the interactions occurring in different exoplanetary systems.

We perform a set of simulations by using stellar wind speed values $\{1, 1.25, 1.5, 2, 2.5, 3, 3.5, 5, 8, 10\}$ V$_{sw}$, stellar wind magnetic field strength values $\{10, 30, 50\}$ nT and planetary dipole field strength values $\{$0.5 {B}$_{e}$, {B}$_{e}$, 2 {B}$_{e}\}$ with all possible combinations between them. Here ${B}_{{e}}$ = $3.1\times 10^4$ nT denotes surface equatorial magnetic field strength of the Earth's dipolar field and V$_{sw}$ is the stellar wind speed. In the present study we assumes a southward-oriented IMF only, the future work will explore the impact of IMF orientation on the magnetospheric configurations.

\begin{table}[htbp]
\centering
\caption{Values of physical parameters used in the simulations and their respective notations.}
\label{table:1}
\begin{tabular}{ l c c } 
 \hline
 \hline
 Physical quantity & Notation & Value used\\ [2ex] \hline
 Density in ambient medium  & $\rho_{amb}$ & 1.5 $\times 10^{-23}$  $\rm{g\thinspace cm^{-3}}$ \\ 
 Pressure in ambient medium & $P_{amb}$ & 2.49 $\times 10^{-11}$ $\rm{ dyne\thinspace cm^{-2}}$ \\
 Density of stellar wind & $\rho_{sw}$ & 4 $\rho_{amb}$\\
 Velocity of stellar wind & $v_{sw}$ & 2.7 $\times 10^{7}$ $\rm{ cm\thinspace s^{-1}}$\\
 Adiabatic index & $\gamma$ & 5/3\\
 Planetary mass & $M_{p}$ & 5.972 $\times 10^{27}$ \rm{g}\\
 Planetary radius & $R_{p}$ & 6.371 $\times 10^8$ \rm{cm}\\
  Intrinsic planetary magnetic field & $B_{p}$ & 0.5$B_{{e}}$, $B_{{e}}$ and  2$B_{{e}}$ $\thinspace^*$\\
 Stellar wind magnetic field strength & $|B_{sw}|$ & 10 nT, 30 nT and 50 nT\\
 Magnetic diffusivity & $\eta$ & 10$^{13}$ $\rm{ cm^{2}\thinspace s^{-1}}$\\
  Magnetospheric tilt angle & $\theta_{p}$ & 11$^\circ$\\
 \hline
\end{tabular}

*\textit{The symbol B$_e$ represents the magnetic field for the case of the Earth's dipole}.
\end{table}

We initialize a conducting plasma atmosphere surrounding the planet, the density profile of which in the vicinity of the planet is defined by
\begin{equation*}
\rho_{{p}} = 10^6 \rho_{{amb}}  \qquad r\leq {R_p}
\end{equation*}
\begin{equation}
\begin{split}
\rho_{{atm}} (r) =  \rho_{{p}} + \frac{(\rho_{{amb}} - \rho_{{p}})}{2} \Big[{\tanh}\Big\{ 9\Big(\frac{r}{R_P} - 2\Big)\Big\}&+1 \Big] \\
{R_P} \leq r \leq {3R_p} \thinspace ,
\end{split}
\end{equation}
where ${\rho_{{p}}}$, ${\rho_{{amb}}}$ and ${\rho_{{atm}}}$  are the densities of the planet, ambient medium and atmosphere respectively and $r$ is the radial distance from the origin. For more details on the justifications for the choice of the above atmospheric profile, please refer to section 2.1 of \cite{basak_nandy_2021}. The pressure distribution in the atmosphere is evaluated by numerical integration of the equation
\begin{equation}
    \frac{{dP}}{{dr}} = -\rho_{{atm}} (r) g(r)
\end{equation}
Here, the gravitational field is given by $g(r) = -\frac{G M_{{p}}}{r^2}$ where $M_{{p}}$ is the planetary mass. The pressure inside the planet is evaluated by extrapolating the value of pressure at the planet-ionosphere boundary. The density and pressure in the region ${r > 3 R_{{p}}}$ is initialized to be equal to that in the ambient medium~(Table~\ref{table:1}).

The input parameters at the stellar wind injection boundary (at $x$= -80 ${R_p}$ in the $yz$ plane) are obtained by solving the Rankine-Hugoniot magnetized jump conditions for the given shock velocity. A southward oriented (SIMF) stellar wind is injected with density 
${\rho}_{{sw}}$ = 4 ${\rho}_{{amb}}$. For all other faces of the Cartesian box, force-free outflow boundary condition is implemented. The parameters defining the properties of the stellar wind and planet are varied in each run and the steady state configuration of the planetary magnetosphere is analyzed.

\section{Results and Discussions}

The nature of interaction between the stellar wind and the planetary magnetosphere is governed by the combined effect of plasma flow speed, stellar and planetary magnetic field strengths, and the resulting Alfvén Mach number. Our three-dimensional magnetohydrodynamic (MHD) simulations explore how variations in stellar wind speed and magnetic field strength shape the global planetary magnetospheric configuration and topology, the Alfvén wing morphology, and associated current systems, especially in the night-side wake region. The results presented here provide insights into the formation and evolution of Alfvén wings, the impact on wing opening angle, and the properties of the magnetotail current sheet under dynamic stellar wind conditions.

As the stellar wind speed increases, the Alfvén Mach number also increases, thereby altering the spatial extent and geometry of Alfvén wings. These findings are directly relevant to solar system planets during extreme solar events and exoplanets orbiting active stars, where such wind conditions occur often \citep{Ridley_2007,Vidotto_2014,Burkholder_2024}.

\subsection{Morphological evolution of Alfvén wings with stellar wind speed}

\begin{figure}[htbp]
    \centering
    \includegraphics[width=1.0\linewidth]{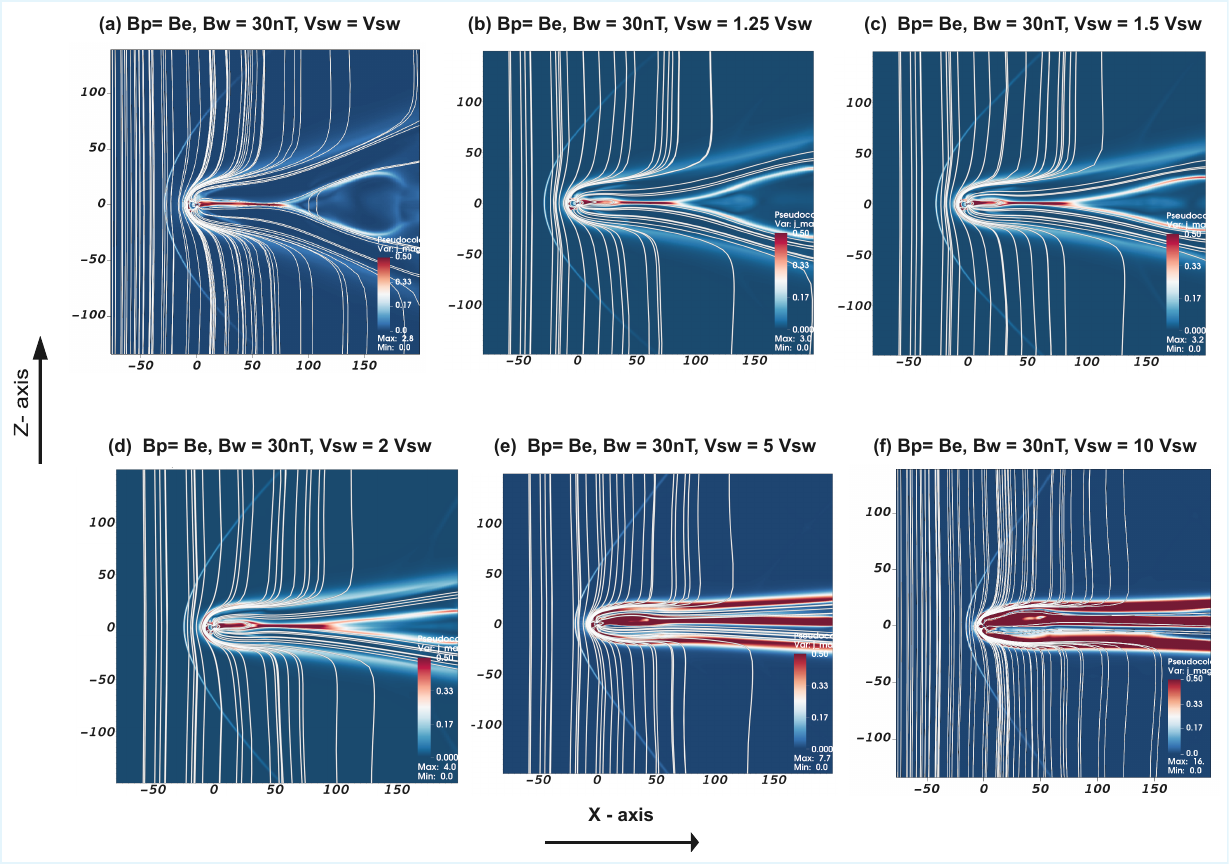}
    \caption{Steady-state planetary magnetospheric configurations for increasing stellar wind speed with the parameters -- planetary dipole field strength $\mathrm{B}_{p} = \mathrm{B}_{e}$ and stellar wind magnetic field $\mathrm{B}_{sw}$ = 30 nT. Panels (a)–(f) illustrate different cases with stellar wind speed ranging from $\mathrm{V} = \mathrm{V}_{sw}$ to 10 $\mathrm{V}_{sw}$. The background colormap shows the magnitude of current density while white streamlines trace the magnetic field lines in the $y=0$ meridional plane. The planet is located at the origin and the stellar wind flows in from the left boundary. Distances are measured in units of the planetary radius.  As the stellar wind speed increases, the day-side magnetospheric compression strengthens due to increasing ram pressure while the night-side magnetotail Alfvén wings slowly close down as evident from the decreasing wing angle.}
    \label{alfven_wing}
\end{figure}

The Alfvén wing geometry in the night-side magnetotail region of the planetary magnetosphere changes considerably as the incoming stellar wind is varied from weak to strong flow conditions. Figure~\ref{alfven_wing} illustrates the morphological evolution of the Alfvén wings in the planetary night-side magnetotail region as a function of the increasing stellar wind speed for fixed values of planetary and stellar wind magnetic fields. The top row [panels (a)–(c)] of the figure \ref{alfven_wing} corresponds to steady-state planetary magnetospheric configuration for stellar wind speed values  V$_{sw}$,  1.25 V$_{sw}$ and 1.5 V$_{sw}$ while the bottom row [panels (d)-(f)] shows the same for wind speeds 2 V$_{sw}$, 5 V$_{sw}$ and 10 V$_{sw}$ respectively. The varying magnetospheric geometry with stellar wind speed as shown in the plots indicate that the ambient plasma flow strongly shapes the planetary environment. When the stellar wind speed is gradually increased keeping other parameters fixed, the Alfvén wings tend to close down slowly which is evident from the decreasing wing opening angle. At the lowest wind speed [Fig.~\ref{alfven_wing}(a)], broad and well‑developed Alfvén wings extend above and below the ecliptic plane in the planetary night-side with a large opening angle forming two nearly symmetric current‑carrying structures that connect the planetary obstacle to the ambient plasma.  This configuration is characterized by strongly 
sub-Alfv\'enic interaction in a low plasma-$\beta$ 
environment. Under these conditions --- where both 
the Alfv\'en Mach number $M_A < 1$ and the plasma 
$\beta \ll 1$ --- perturbations excited at the 
planet propagate preferentially along the magnetic field direction, efficiently channeling energy into the Alfv\'en wings and inflating them laterally. In the magnetotail region downstream of the planet, the 
magnetic pressure dominates over the thermal 
pressure, ensuring that the low-$\beta$ condition is satisfied and the wing structure is maintained. For the highest wind speed however [Fig.~\ref{alfven_wing}(f)], the wings in the night-side cannot be distinguished separately since they come close together with negligible opening angle, straighten out and become almost parallel to the ecliptic plane. The plasma flow velocity is now strong enough to generate ample ram pressure that forces the wings down and aligns the extended magnetosphere along the flow direction. This increased ram pressure is also responsible for compressing the planetary field at the day-side leading to a lower magnetopause stand-off distance. This will be discussed in greater detail in the upcoming paragraphs. The overall trend of the plots in the above figure reflects that as the Alfvén Mach number (M$_A$) increases, the ability of the plasma to communicate disturbances across field lines diminishes, confining Alfvénic perturbations more tightly to the direction of the flow. In our simulations, this progressive “closure” of the wings from wide open structures at low V$_{sw}$ to more collimated, flow‑aligned wings at higher V$_{sw}$, is evident across the panels (a)–(f). The results also reveal that the current density is highest along the wing boundaries, indicating that these structures are defined by intense field-aligned currents.

\begin{figure}[htbp]
    \centering
    \includegraphics[width=1\linewidth]{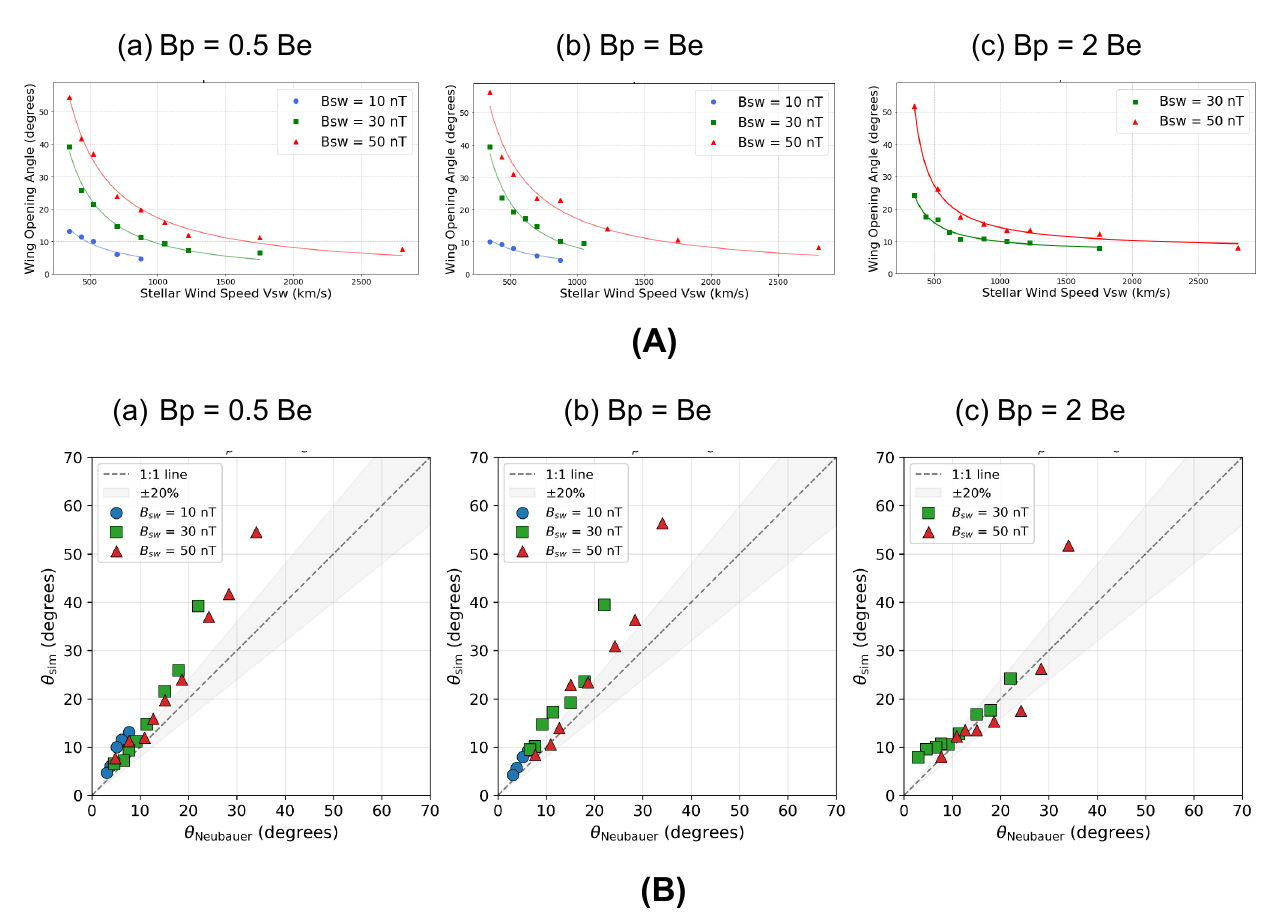}
    \caption{ (A) Plots of the night-side Alfvén wing opening angle as a function of stellar wind speed for different stellar and planetary magnetic field strengths. Panels (a), (b), and (c) correspond to planetary dipole field strengths of 0.5 $\mathrm{B}e$, 1.0 $\mathrm{B}e$, and 2.0 $\mathrm{B}e$ respectively. In each panel, the data points in blue, red, and green colors are for stellar wind magnetic field 10 nT, 30 nT and 50 nT respectively while the solid curves show the corresponding best-fit trends to these simulation data. 
    (B) Comparison of simulated Alfvén wing opening angles ($\theta_{\mathrm{sim}}$) with the analytical prediction of \citep{Neubauer_1980}, $\theta_{\mathrm{Neubauer}} = \arctan(1/M_A)$. The dashed line indicates perfect agreement and the shaded band marks $\pm$20\% deviation. The plots illustrate that for a fixed planetary field, the Alfvén wings open up for a higher stellar wind magnetic field due to a stronger force acting in the magnetotail region. However, when the planetary field strength is increased, it opposes this force and tends to close the wings down again. The interplay between stellar and planetary magnetic fields in addition to stellar wind speed is responsible for the Alfvén wing geometry and the steady-state magnetospheric topology.
    }
    \label{opening_angle}
\end{figure}

In order to provide a proper quantification of the morphological trends observed in Fig.~\ref{alfven_wing}, we plot the Alfvén wing opening angle as a function of both stellar wind speed and magnetic field for three values of planetary dipole field. In Fig.~\ref{opening_angle}(A), the panels (a), (b) and (c) correspond to a given planetary field strength, i.e. $\mathrm{B}p$ = 0.5 $\mathrm{B}e$, $\mathrm{B}e$ and 2 $\mathrm{B}e$ respectively. In each panel, blue, green and red dots show the measured opening angles for simulation data with B${sw}$ = 10 nT, 30 nT and 50 nT respectively as a function of V${sw}$. The corresponding smooth curves represent best fits to the simulation data. Let us first consider the scenario for a fixed planetary field. In any of the given panels in Fig.~\ref{opening_angle}(A), it is observed that for slow winds where the interaction remains strongly sub‑Alfvénic, the wing opening angle is larger. A stronger planetary dipole creates a more "rigid" obstacle that resists the lateral inflation of the wings, forcing them to remain closer to the ecliptic plane. As $\mathrm{V}{sw}$ is increased keeping $\mathrm{B}{sw}$ constant, the wing opening angle gradually decreases which is the effect of the increasing ram pressure of the wind plasma that shuts the wings down and aligns them in its direction. When $\mathrm{V}{sw}$ approaches the super-Alfvénic limit, the wing opening angle reaches an asymptotic minimum -- this indicates the extent up to which the wings may be closed for a given value of $\mathrm{B}{sw}$. The results depict a regime change where the kinetic energy of the wind completely dominates the magnetic tension of the wings, effectively 'pinching' the structures into a tail-like geometry.

Considering the influence of varying magnetic fields on the Alfvén wing topology as shown in Fig.~\ref{opening_angle}(A) we observe that for a given stellar wind speed V${sw}$ and planetary dipole field B${p}$, as the stellar wind magnetic field strength B${sw}$ is increased from 10 nT to 50 nT, the wing opening angle becomes larger for a higher value of B${sw}$ in all three panels. A higher value of B${sw}$ indicates a larger local Alfvén speed and the tendency for the interaction to remain more sub‑Alfvénic over a wider range of V${sw}$. A stronger stellar wind magnetic field pulls the wings outwards away from the ecliptic plane due to the existence of a stronger Lorentz force acting on the plasma field lines in the magnetotail region -- this increases the wing opening angle. Conversely, increasing the planetary magnetic field across the three panels [Fig.~\ref{opening_angle}(A) (a)-(c)] reduces the wing opening angle for fixed values of V${sw}$ and B${sw}$. This implies that a stronger magnetic obstacle confines the wings more tightly around the magnetosphere and requires stronger wind forcing to open them to comparable extents. This occurs because the Lorentz force due to a stronger planetary field acts in a direction opposite to that due to the stellar magnetic field and tends to close the wings down. It is pertinent to mention here that in Fig.~\ref{opening_angle}(A)(c), we show the wing opening angle only for stellar wind magnetic field $\mathrm{B}{sw}$= 30 nT and 50 nT because for the case $\mathrm{B}{sw}$ = 10 nT, the Alfvén wings do not open sufficiently due to a very strong planetary field as compared to the stellar wind magnetic field and therefore, the wing opening angle is negligible for this case.

To validate the Alfvén wing physics captured by our simulations, we compare the measured wing opening angles with the analytical prediction of \citet{Neubauer_1980}, $\theta_A = \arctan(1/M_A)$, which assumes a point-source conducting obstacle in a uniform magnetized plasma flow. Fig.~\ref{opening_angle}(B) shows the simulated opening angles ($\theta_{\mathrm{sim}}$) plotted against the corresponding Neubauer predictions ($\theta_{\mathrm{Neubauer}}$) for all parameter combinations across the three planetary dipole strengths. The data follow a clear positive correlation with the analytical values across all panels, confirming that the Alfvén Mach number is the primary controlling parameter for the wing geometry. Few data points for the simulated angles exceed the Neubauer prediction with data points lying above the 1:1 line. This departure arises because the Neubauer formulation treats the obstacle as a point conductor with zero spatial extent, whereas in our simulations the planetary magnetosphere acts as a finite, inflated obstacle that deflects the incoming stellar wind over a cross-section significantly larger than the planetary radius. This effectively broadens the interaction region and leads to the formation of the wing current sheets at larger angular separations than predicted by the point-source model. 

The above results provide a comprehensive characterization of Alfvén wing geometry across the three-dimensional parameter space of stellar wind speed, stellar magnetic field, and planetary dipole strength. The validated agreement with the \citet{Neubauer_1980}, along with the quantified departures arising from finite magnetospheric effects, provides a foundation for interpreting observed wing signatures to underlying stellar wind conditions in both the solar system and (exo)planetary environments.

\subsection{Magnetotail current sheet response to magnetopause variability}

The response of the planetary magnetotail to changing stellar wind conditions is closely related to the behaviour of the day-side magnetopause. We have seen that an increase in stellar wind ram pressure compresses the day-side magnetopause and alters the global magnetospheric configuration, the geometry of the Alfvén wings and the structure of the magnetotail current sheet. In order to quantify this relationship, we analyze the night-side magnetotail current sheet length ($J_{cs}$) variation as a function of the day-side magnetopause stand-off distance ($R_{mp}$) for a range of stellar wind speeds and magnetic field strengths for different planetary dipole field strengths. For studying the night-side response, we define the magnetotail current sheet length ($J_{\mathrm{mag}}$) as the distance from the planet to the point where the current sheet begins to bifurcate in the magnetotail wake region. For details on the identification of the bifurcation region and the computation of current sheet length, kindly refer the Results and Discussion section of \citep{Gupta_2023}.

\begin{figure}
    \centering
    \includegraphics[width=1\linewidth]{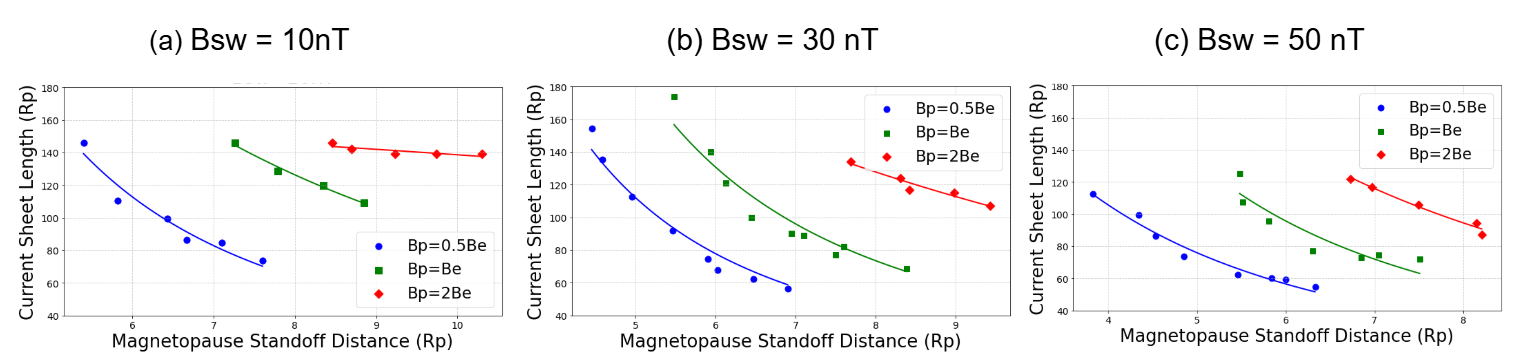}
    \caption{Relationship between the night-side magnetotail current sheet length and the subsolar magnetopause standoff distance for different planetary and stellar wind magnetic field strengths. Panels (a), (b) and (c) correspond to stellar wind magnetic field 10 nT, 30 nT and 50 nT respectively. In each panel, data points in blue, green and red colors correspond to planetary dipole strengths of 0.5 $\mathrm{B}_{e}$, 1.0 $\mathrm{B}_{e}$, and 2.0 $\mathrm{B}_{e}$ respectively while the solid curves show the corresponding best-fit to the simulation data. As expected, the night-side current sheet length and the day-side magnetopause stand-off distance show an inverse relation between them for fixed parameter values. Increasing the planetary field keeping stellar field constant increases the values of both current sheet length and magnetopause stand-off distance while increasing the stellar field keeping planetary field constant lowers their values.}
    \label{jmag_Rmp}
\end{figure}

Figure~\ref{jmag_Rmp} shows the variation of magnetotail current sheet length with day-side magnetopause stand-off distance $R_{mp}$ for three values of stellar wind magnetic field B$_{sw}$ = 10 nT, 30 nT and 50 nT in panels (a), (b) and (c) respectively. In each panel, the simulation data points and their corresponding best-fit curves are shown for  planetary magnetic field strengths $B_p = 0.5 B_e$, $B_p = B_e$, and $B_p = 2B_e$ in blue, green and red colors respectively. The data points for any given curve are obtained by varying the stellar wind speed and measuring the corresponding magnetopause stand-off distance and magnetotail current sheet length for a particular wind speed. We first consider a single case, for instance the blue curve in panel (a), for understanding the trend of the results so obtained. The following explanation however holds for any given case in the present figure. The data exhibit a clear inverse relationship between magnetopause stand-off disance and the current sheet length, well captured by the fitted curves (see Fig.~\ref{jmag_Rmp}). As R$_{mp}$ increases, the magnetotail current sheet becomes shorter and vice-versa. When the stellar wind speed is lower, the ram pressure is also less and so is the day-side magnetospheric compression which leads to a larger magnetopause stand-off distance. When the wind speed is gradually increased, the ram pressure increases which compresses the day-side magnetosphere even more and lowers the value of the stand-off distance. The higher wind speed advects the night-side magnetotail and stretches it further along its direction which results in a longer current sheet in the wake region. This indicates that as the wind speed increases, the day-side penetration is higher resulting in a lower stand-off distance while the night-side advection leads to a longer current sheet. This behavior reflects the response of the magnetosphere to enhanced stellar wind forcing where the day-side and night-side properties are inversely related to each other. The elongation of the current sheet observed here is intrinsically linked to the 'closure' of the Alfvén wings discussed in section 3.1 -- as the wings align more closely with the flow, the magnetosphere is transformed into a more stretched, tail-dominated geometry. 

Now, let us consider any of the three panels in Fig.~\ref{jmag_Rmp} for understanding the impact of varying planetary magnetic field on the magnetospheric topology. We find that as the planetary field is increased (from blue to red via green), the curves are shifted to larger values of both day-side stand-off distance and night-side current sheet length. For a stronger planetary magnetic field, the day-side magnetic pressure close to the planet is higher which nullifies the stellar wind ram pressure further away from the planetary surface resulting in a larger value of magnetopause stand-off distance. A stronger intrinsic dipole field increases the internal magnetic pressure which effectively 'inflates' the entire magnetospheric cavity. We have already seen in Fig.~\ref{opening_angle} that a stronger planetary magnetic field tends to close the Alfvén wings down in the magnetotail region as evident from the decreasing wing opening angle. The more the wings come together, the longer is the magnetotail current sheet due to oppositely directed magnetic field lines in the upper and lower halves of the ecliptic plane. Therefore, a stronger planetary field results in a longer night-side current sheet.

For understanding the impact of varying stellar wind magnetic field on the magnetospheric structure, let us move across the panels from left to right in Fig.~\ref{jmag_Rmp} for increasing $B_{sw}$. When the stellar wind magnetic field is low, magnetic reconnections at the planetary day-side are less which leads to a larger magnetopause stand-off distance. When the stellar magnetic field considered in this study is increased, the reconnection rate increases and the stellar penetration is more into the planetary magnetosphere thereby decreasing the stand-off distance. For understanding the dynamics in the night-side, we again refer to Fig.~\ref{opening_angle} where it is found that a stronger stellar field pulls the Alfvén wings apart due to higher magnetic tension as depicted by the larger wing opening angle -- this geometry ensures that only a small portion of the oppositely directed field lines are close to each other near the ecliptic plane and this leads to a shorter magnetotail current sheet. This suggests that a strongly magnetized stellar wind acts as a confining sheath that limits the expansion of the magnetotail regardless of the day-side compression state. 

This inverse relationship is consistently observed across all stellar magnetic field cases shown in Fig \ref{jmag_Rmp} (a)–(c), indicating that the coupled dayside–nightside response is robust across the explored parameter space.Summarizing the impact of varying stellar wind speed and stellar and planetary magnetic fields on the magnetospheric geometry, we note that a higher V$_{sw}$ compresses the magnetosphere on the day-side and elongates it on the night-side, a stronger B$_{sw}$ increases magnetic confinement, leading to a more compact magnetosphere and shorter magnetotail current sheet, and a stronger B$_p$ leads to a more extended magnetosphere in both directions.

\subsection{Properties of plasma flow and magnetic field inside the Alfvén wings}

For understanding the stellar wind crafted planetary environment for different parameter values, we analyze the local plasma properties along the $z$-axis in the in the y = 0 meridional plane at a fixed downstream location x = 100 R$_p$ in the night-side magnetotail.

\begin{figure}
    \centering
    \includegraphics[width=1\linewidth]{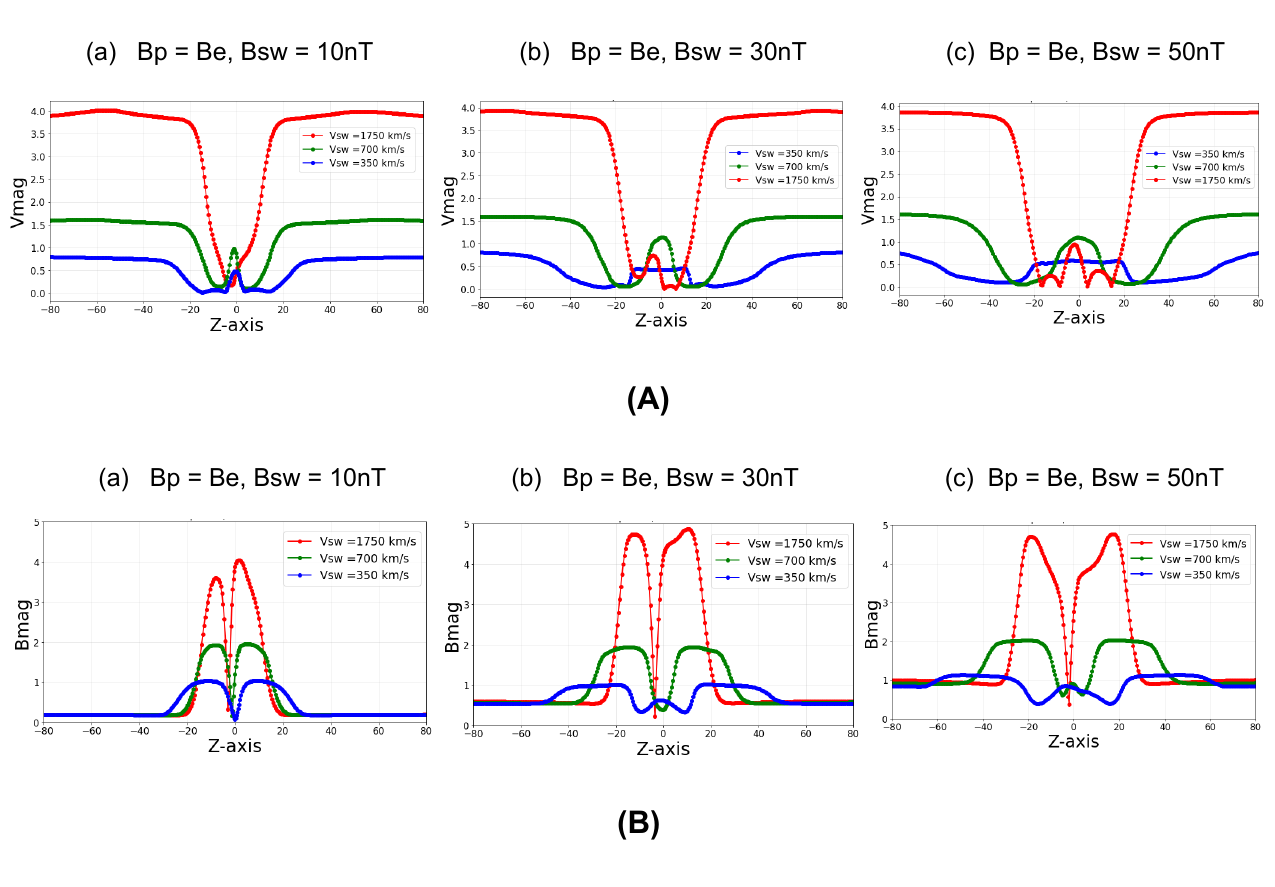}
    \caption{Plasma flow velocity and magnetic field strength inside the Alfvén wings for $\mathrm{B}_p = \mathrm{B}_{e}$ with varying stellar wind conditions. Panel (A) shows the flow velocity magnitude and panel (B) shows the magnetic field magnitude inside the wings. Blue, green and red curves correspond to stellar wind speeds of $\mathrm{V}_{sw}$, 2 $\mathrm{V}_{sw}$, and 5 $\mathrm{V}_{sw}$ respectively. In each panel, sub-panels (a), (b) and (c) represent cases with stellar wind magnetic field strengths of 10 nT, 30 nT and 50 nT respectively. As expected, magnetic field lines are clustered inside the night-side Alfvén wings leading to magnetic field accumulation while the plasma flow significantly decreases inside the wings in the magnetotail region.} 
    \label{fig:v_b_inside_wings}
\end{figure}

Figure~\ref{fig:v_b_inside_wings} (A) depicts the variation of plasma velocity (V$_{mag}$) as a function of the distance along $z$-axis from $-80 R_p$ to $80 R_p$ for $B_p = B_e$ and three different stellar wind speeds (V$_{sw}$, 2 V$_{sw}$ and 5 V$_{sw}$) shown in blue, green and red colors respectively. The stellar wind magnetic field strength increases across the panels (a)-(c) from 10 nT to 50 nT. In all cases, there is a marked reduction in flow speed within the wings compared to the ambient medium i.e. the flow decelerates inside the Alfvén wings. If we move across panels (a)-(c) in Fig.~\ref{fig:v_b_inside_wings}(A), the lateral extent of the Alfvén wings increases with stellar wind magnetic field strength whereas for any fixed combination of planetary and stellar magnetic fields, the wings become narrower when the stellar wind speed is higher. The red curves ($5 V_{sw}$) show the most prominent drop in velocity. This suggests that as the stellar wind speed increases, the Alfvén wings become regions of intense drag, where a larger fraction of the wind’s kinetic energy is converted into magnetic energy (and potentially thermal energy through dissipation processes), leading to enhanced magnetic pressure within the wings.

The corresponding magnetic field magnitude profiles are plotted in Fig.~\ref{fig:v_b_inside_wings}(B) along the same route. The plots show that the field strength inside the wings is consistently higher than the background ambient magnetic field. This "pile-up" of magnetic field occurs due to the clustering of planetary magnetic field lines reconnected with the stellar field in both the upper and lower parts of the magnetotail. The orientation of these two structures however depends on the parameter values, i.e. stellar wind speed and stellar and planetary magnetic fields.
The enhanced magnetic pressure within the wings opposes the incoming plasma flow, leading to a significant reduction in velocity. For all three panels (a)-(c) in Fig.~\ref{fig:v_b_inside_wings}(B), it is evident that higher incoming stellar wind speeds are associated with larger magnetic field magnitudes inside the wings, implying that faster flows advect the magnetotail and compress the magnetic flux more. In addition, when the stellar wind magnetic field is increased, the regions of enhanced magnetic field broaden, indicating that stronger upstream fields produce wider Alfvén wings.

These results demonstrate that Alfvén wings are not merely geometric structures but regions of significant dynamical interaction, where plasma flow is slowed and magnetic flux is enhanced. These dynamics plays a key role in mediating energy transfer between the stellar wind and the planetary magnetosphere.

\subsection{Magnetotail current sheet as a function of stellar wind properties}

While previous sections characterize the morphology and internal properties of Alfvén wings, we now investigate the global response of the magnetotail to combined changes in stellar wind forcing -- a linear scaling between the night-side magnetotail current sheet length ($J_{cs}$) and the composite parameter $V_{sw}^2 / B_{sw}$. This choice is physically motivated by magnetotail currentsheet length scalings in which the currentsheet length increases with increasing stellar wind dynamic pressure ($P_{d} \propto V_{sw}^2$) (ref fig(\ref{alfven_wing}) and decreases with increasing interplanetary magnetic field strength \citep{Gupta_2023}. The ratio between the dynamic pressure of the stellar wind (proportional to V$_{sw}^2$) and the magnetic confinement exerted by the stellar wind (proportional to B$_{sw}$), effectively measures the relative importance of plasma kinetic pressure to the confining stellar wind magnetic field strength.

\begin{figure}
    \centering
    \includegraphics[width=1\linewidth]{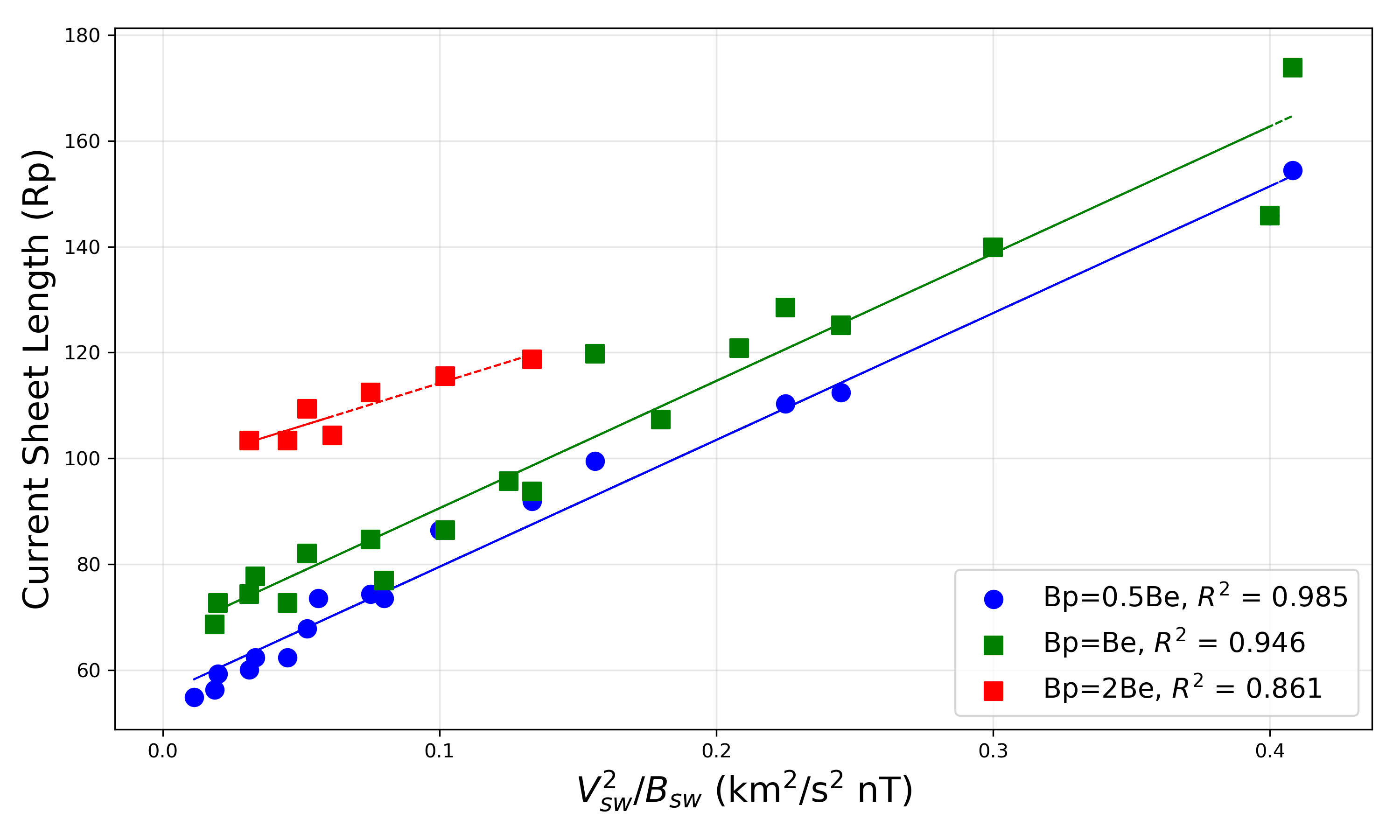}
    \caption{Night-side magnetotail current sheet length as a function of the composite parameter $\mathrm{V}_{sw}^2 / \mathrm{B}_{sw}$ for different planetary dipole field strengths where Alfvén wings form and finite current sheet lengths can be computed from simulations. Blue, green and red curves correspond to $\mathrm{B}_p$ = 0.5 $\mathrm{B}_e$ , $\mathrm{B}_e$ and 2 $\mathrm{B}_e$ respectively. Each curve aggregates simulation data across the stellar wind magnetic field strengths 10 nT, 30 nT and 50 nT.}
    \label{Jmag_v2_vsw}
\end{figure}

Figure~\ref{Jmag_v2_vsw} shows the night-side magnetotail current sheet length as a function of V$_{sw}^2$/B$_{sw}$. The simulated data points are well described by a linear relation for all three planetary dipole strengths, with coefficients of determination R$^2$ =0.985, 0.946, and 0.861 for B$_p$=0.5 B$_e$, B$_e$ and 2B$_e$ respectively, confirming the robustness of the scaling. This linearity holds across a wide range of $M_A$ regimes, suggesting that the stretching of the magnetotail behave predictably in response to the kinetic driving and magnetic tension. For all the cases, while the linear trend is consistent, the slope and intercept are determined by the planet's intrinsic magnetic field. The blue ($0.5 B_e$), green ($1.0 B_e$), and red ($2.0 B_e$) curves show that for any given stellar wind condition, a stronger planetary dipole consistently maintains a longer current sheet.

This behavior indicates that stronger wind driving relative to the ambient stellar wind magnetic field is associated with a more extended magnetotail current sheet, whereas stronger B$_{sw}$ at fixed V$_{sw}$ is associated with a more compact tail. Within the range of parameters explored, this provides an empirical scaling that links upstream stellar wind conditions to the size of the magnetotail current sheet. The slope for current sheet length B$_{p}$ = 2B$_e$ is less due to the fact that V$_{sw}^2$/B$_{sw}$ will lead to less disruption in the current sheet length with respect to B$_{p}$ = 0.5 B$_e$ and B$_e$. The scaling remains valid as the system transitions from the sub-Alfvénic regimes (where Alfvén wings are prominent) toward more Earth-like super-Alfvénic conditions. This implies that the $V_{sw}^2 / B_{sw}$ parameter is robust to describe the magnetotail stretching across diverse star-planet systems.

\section{Conclusion}

The diversity of star-planet interactions that occur in the cosmos influence the dynamics of different planetary systems. In this work, we have explored several such (exo)planetary systems by carrying out a vast parameter-space study using 3D global magnetohydrodynamic (MHD) simulations -- incorporating variations in stellar wind speed ($V_{sw}$), stellar magnetic field ($B_{sw}$), and  planetary magnetic field ($B_p$) strengths. The first two parameters essentially represent the stellar wind forcing, while the last one characterizes the planetary magnetosphere. Variations in the above parameters influence the steady-state planetary environment and topology, especially the day-side magnetopause and the night-side Alfvén wings along with the associated magnetotail current sheets. By investigating a wide range of upstream conditions ranging from sub-Alfvénic to moderately super-Alfvénic flows, our parameter-space study -- comprising ninety simulations -- aims to illuminate the star-planet interaction physics at play and the consequent magnetospheric response.

The Alfvén wings emerge prominently under sub-Alfvénic conditions and show an inverse scaling of the wing opening angle with stellar wind speed -- lower wind speeds lead to broad, open wings, while higher speeds yield narrow, elongated structures. This morphological evolution reflects that increasing Alfvén Mach number compresses the wings to a more collimated structure along the flow direction, When the stellar wind magnetic field is increased keeping other parameters fixed, it tends to pull the wings apart resulting in a larger wing opening angle. A stronger planetary field, on the other hand, opposes the Lorentz force of the stellar wind and tends to close the wings down by bringing them closer to the ecliptic plane. The relative strength of these three parameters is, therefore, responsible for the steady-state morphology of the wings and the planetary environment. We further validate the simulated Alfvén wing opening angles against the analytical prediction of Neubauer (1980) with slight deviations arising from finite magnetospheric size.

Our analysis also reveals an inverse relationship between the subsolar magnetopause stand-off distance and the night-side magnetotail current sheet length for different values of the stellar wind speed. This trend can be explained using pressure balance -- as the wind speed is increased, the ram pressure increases, which gradually compresses the day-side magnetosphere and stretches out the night-side magnetotail --  resulting in a lower magnetopause stand-off distance and a longer magnetotail current sheet, respectively. A stronger stellar wind field compresses the magnetosphere, while a stronger planetary field expands the magnetosphere from both ends. We analyse the plasma flow inside the Alfven wings. Within the wings, plasma flow decelerates while magnetic field strength enhances. The lateral extent of Alfvén wing increases with B$_{sw}$ while with increasing stellar wind speed the wings become narrower. We also analyze the impact of the composite parameter V$_{sw}^2$/B$_{sw}$, which is a proxy for the the ratio of dynamic pressure to stellar magnetic field strength (associated with the magnetotail current sheet length). The current sheet length scales approximately linearly with this parameter, suggesting a relation in which a stronger net kinetic driving relative to the magnetic confinement elongates the tail. 

Our study presents a framework that relates stellar wind speed, stellar magnetic field, and planetary magnetic properties to the global evolution of magnetospheric morphology, Alfvén wing dynamics and magnetotail current sheets. The results indicate the relevance to solar system planets encountering extreme solar events, such as coronal mass ejections that reduce the Alfvén Mach number and trigger wing formation, as well as (exo)planets orbiting active stars where sub-Alfvénic conditions prevail routinely. Our study is relevant for understanding how sub- and super-Alfvénic wind regimes influence the magnetospheric topology in both solar and extrasolar environments. By establishing qualitative scalings for wing morphology, tail structure, and plasma properties, our simulations provide a foundation for interpreting the observations of Alfvén wing structures and associated magnetospheric characteristics. We expect that these insights will motivate and inform future exploration of star-planet interactions in diverse astrophysical contexts.

\section*{Acknowledgments}
The development of the Star - Planet Interaction module (CESSI-SPIM) and the simulations were carried out at the Center of Excellence in Space Sciences India (CESSI), supported by IISER Kolkata, Ministry of Education, Government of India. We gratefully acknowledge the Santimay and Sunanda Basu Foundation for enabling the computational infrastructure used in this study. SG acknowledges Dr. Jagannath Nayak, Director CHESS, and Saumendra Nath Datta, Technology Director at CHESS, for the necessary approval and encouragement to continue her PhD research work at CESSI.

\bibliographystyle{Frontiers-Harvard} 
\bibliography{manuscript}

\end{document}